
\magnification=\magstephalf
\hsize=5.5truein
\vsize=7.4truein
\hoffset=0.5truein
\font\lgbf=cmbx10 scaled\magstephalf
\font\smrm=cmr10 at 10truept
\font\smcap=cmcsc10
\font\tenmib=cmmib10
\newfam\mibfam \def\mib{\fam\mibfam\tenmib} \textfont\mibfam\tenmib
\mathchardef\zeta="7110 \mathchardef\xi="7118
\def\half{{\raise1pt\hbox{$\scriptscriptstyle 1/2$}}}
\def\neghalf{{\raise1pt\hbox{$\scriptscriptstyle -\!1/2$}}}
\def\dS{de\thinspace Sitter}

\def\adS{anti--\dS}
\def\SadS{Schwarzschild--\adS}

\def\Real{\rm I\kern-0.2em R}
\def\sub#1{{\lower1pt\hbox{$\scriptstyle #1$}}}
\newcount\secnum
\def\clearsecnum{\secnum=0} \clearsecnum
\newcount\eqcount
\def\cleareqcount{\eqcount=0}
\def\eqnum{(\the\secnum.\the\eqcount)}
\def\eq{\global\advance\eqcount by 1 \eqnum}
\outer\def\section#1\par{\bigbreak\advance\secnum by 1
  \cleareqcount\centerline{\smcap\romannumeral\the\secnum. #1}
  \nobreak\medskip\noindent}

\def\B3{{\cal B}}
\def\dDx{{d^D x}}
\def\dDmx{{d^{D-1}x}}
\def\dDmmx{{d^{D-2}x}}
\def\Pres{{{\cal P}}}

\def\({{\scriptscriptstyle (}}
\def\){{\scriptscriptstyle )}}
\def\sss0{{\scriptscriptstyle 0}}
\def\sssH{{\scriptscriptstyle H}}
\vglue 0.5truein

\centerline{\hfill WATPHYS TH-94/02}
\centerline{\hfill CTMP/006/NCSU}
\vskip 1cm
\centerline{\lgbf Temperature, Energy, and Heat Capacity of}
\smallskip
\centerline{\lgbf Asymptotically Anti--de Sitter Black Holes}
\vglue 1cm
\centerline{\bf J.D. Brown\footnote*{\smrm Departments of Physics and
Mathematics, North Carolina State University, Raleigh, NC 27695--8202},
J. Creighton\footnote{$^{\dag}$}{\smrm Department of Physics, University
of Waterloo, Waterloo, Ontario N2L 3G1} and R.B. Mann{$^{\dag}$} }
\vglue 1cm


{\narrower\noindent\smrm

We investigate the thermodynamical properties of black holes in (3+1)
and (2+1) dimensional Einstein gravity with a negative cosmological
constant. In each case, the thermodynamic internal energy is computed
for a finite spatial region that contains the black hole. The
temperature at the boundary of this region is defined by differentiating
the energy with respect to entropy, and is equal to the product of the
surface gravity (divided by~$2\pi$) and the Tolman redshift factor for
temperature in a stationary gravitational field. We also compute the
thermodynamic surface pressure and, in the case of the (2+1) black hole,
show that the chemical potential conjugate to angular momentum is equal
to the proper angular velocity of the black hole with respect to observers
who are at rest in the stationary time slices. In (3+1) dimensions, a
calculation of the heat capacity reveals the existence of a
thermodynamically stable black hole solution and a negative heat capacity
instanton. This result holds in the limit that the spatial boundary
tends to infinity only if the comological constant is negative; if the
cosmological constant vanishes, the stable black hole solution is
lost. In (2+1) dimensions, a calculation of the heat capacity
reveals the existence of a thermodynamically stable black hole solution,
but no negative heat capacity instanton.
\par}


\section introduction

A variety of theoretical arguments indicate that black holes have
thermodynamical properties. This thermal character is expected to
hold for all black holes, yet much of the literature on black hole
thermodynamics is restricted to the case of spacetimes that are
asymptotically flat in spacelike directions. Since asymptotic
flatness is not always an appropriate theoretical idealization,
and is never satisfied in reality, it is important to develop a
theoretical framework for the description of black hole
thermodynamics that is divorced from the assumption of asymptotic
flatness of the spacetime. This is one of the primary motivations behind the
formalism developed in Refs.~[1--8]. This approach to black hole
thermodynamics can be applied to gravitational and matter
fields within a bounded, finite spatial region, so the
asymptotic behavior of the gravitational field becomes
irrelevant. In this way, it is possible to treat black hole spacetimes
that are asymptotically curved and black holes in spatially
closed universes. Even
for black hole spacetimes that are asymptotically flat, there are several
advantages to be gained by working in a spatially
finite region~[8]. For example, with the temperature fixed at
infinity, the heat capacity for a Schwarzschild black hole is
negative~[9] and the formal expression for the partition
function is not logically consistent~[10].
On the other hand, with the temperature fixed at a finite
spatial boundary, the heat capacity is positive and there is
no inconsistency in the black hole partition function~[1].

In this paper we employ spatially finite boundary conditions to
investigate the thermodynamical properties of black holes in
(3+1) and (2+1) dimensional Einstein gravity with a negative
cosmological constant. In these cases, the spacetimes are not
asymptotically flat but, rather, they are asymptotically \adS.
Previous calculations of the thermodynamics of asymptotically
\adS~black holes have not been completely correct, because the
temperature is identified with~$\kappa_\sssH/(2\pi)$ where
$\kappa_\sssH$ is the surface gravity of the black hole and
the thermodynamic internal energy is identified with the
conserved mass at infinity~[11,12]. However, the surface
gravity and the mass at infinity each depend on the
normalization of a timelike Killing vector field, and in the
absence of an asymptotically flat region there is no
physically preferred choice. Moreover, for asymptotically
\adS~spacetimes, the physical temperature of a black hole
or any hot object should redshift to zero at spatial
infinity~[13].

We begin in Sec.~2 with a review of Ref.~[7]. This includes
a definition of the total energy~$E$ of the gravitational
field within a region of space with boundary~$B$. In addition,
conserved charges are defined whenever the history of the
boundary~$B$ admits a Killing vector field. The charge
associated with a rotational symmetry is the angular
momentum~$J$, and the charge associated with a timelike
Killing vector field defines a conserved mass~$M$. The
energy~$E$ and the mass~$M$ are not the same---we
discuss the distinction between them. We also show that
the conserved charges as defined here, in the limit that
the boundary~$B$ is pushed to spatial infinity, agree with
the ADM charges~[14] that are defined through an analysis
of the surface terms in the gravitational Hamiltonian~[15].
A spatial stress tensor~$s^{ab}$ for the boundary~$B$
is also defined.

There is some freedom of choice in the definitions of $E$,
$M$, and $J$, which is reflected in the presence
of two arbitrary functions of the boundary geometry. This
choice determines the `zero--point configuration', that is,
the gravitational canonical data for which $E$, $M$, and~$J$
vanish. The two arbitrary functions can be defined
so that $E$, $M$, and $J$ vanish for a stationary
slice of \adS~spacetime, or for a stationary slice of flat
spacetime. (For the (2+1) case, the zero point configuration
can be chosen as a stationary slice of the zero--mass black
hole solution.)

In Sec.~3 we compute the energy  and spatial stress for
the region of a static slice of \SadS~spacetime within a
spherical boundary~$B$. We then identify~$E$ as the
thermodynamic internal energy and the entropy~$\cal S$ as one-quarter
the area of the black hole event horizon. The
temperature~$T$ at the boundary~$B$ is defined by
$\partial E/\partial {\cal S}$, and is equal to the product
of~$\kappa_\sssH/(2\pi)$ and the redshift factor~[13] for temperature
in a stationary gravitational field. In particular, the temperature
depends on the location of the boundary~$B$, and correctly redshifts to
zero in the limit~$B\to\infty$. Moreover, the temperature is
independent of the choice of a zero point configuration for
the energy. We also find that the surface
pressure~$\Pres$, as defined by the derivative of~$E$ with respect
to the area of the boundary~$B$, is given by the trace of the
spatial stress tensor~$s^{ab}$.

Our results show that there are no \SadS~black hole solutions with
temperature at~$B$ less than some critical value~$T_0$, and
there are two possible black hole solutions with a given temperature~$T$
greater than~$T_0$. Moreover, for~$T>T_0$, the  smaller of the two black
holes has a negative heat capacity and the larger of the two
black holes has a positive heat capacity. These results hold in
the limit of a vanishing cosmological constant, and are
interpreted as follows~[1,11]. For low temperatures, the
equilibrium states are described semiclassically by thermal
gravitons propogating on flat or \adS~backgrounds. For
high temperatures, the equilibrium states are classically
approximated by the larger black hole with positive
heat capacity. The Euclidean section of the smaller black
hole is an instanton that dominates the semiclassical
evaluation of the rate of nucleation of black holes~[16]
from flat or \adS~space.

We analyze solutions in the limit in which the boundary goes to
infinity and the temperature is adjusted so that the
black hole horizon size remains fixed. The results depend
crucially on whether the cosmological constant is strictly
negative or zero. If the cosmological constant is negative,
then our results are qualitatively unchanged in this limit.
In particular, solutions include both a large, thermodynamically stable
black hole and a small black hole instanton. If the cosmological
constant is zero, the large black hole is lost in the limit
and only the small black hole instanton solution remains. This shows
that a black hole in infinite space can be thermodynamically stable if the
cosmological constant is negative, but not if the cosmological
constant is zero~[11].

In Sec.~3 we also compute the conserved mass~$M$ for the
\SadS~black hole. (The angular momentum vanishes.) With
an appropriate choice of zero point configuration, and
in the limit~$B\to\infty$, $M$~is equal to the black hole mass
parameter and $\partial M/\partial {\cal S}$~is equal
to~$\kappa_\sssH/(2\pi)$. As mentioned above,
$\kappa_\sssH/(2\pi)$~is not the physical temperature at
infinity. However, in the limit that the boundary tends to
infinity and the black hole horizon size remains fixed,
$\partial M/\partial(\kappa_\sssH/2\pi)$~yields the correct
expression for the heat capacity.

In Sec.~4 we analyze the thermodynamical properties of the stationary
black hole solution to (2+1) dimensional Einstein gravity with a
negative cosmological constant~[12,17]. In this case, there are no
black hole solutions in the limit that the cosmological constant vanishes.
First we compute the energy, spatial stress, mass, and angular
momentum for a stationary slice of
the (2+1) black hole spacetime within an axially symmetric boundary.
The entropy~${\cal S}$ is twice the `area' (circumference) of the
event horizon, and we again identify~$E$ as the thermodynamic
internal energy. Just as for the \SadS~black hole in (3+1) dimensions,
the temperature at the boundary~$B$ is the product of~$\kappa_\sssH/(2\pi)$
with the redshift factor, and the surface pressure is given by the
trace of the spatial stress. The chemical potential conjugate to
angular momentum is defined by~$\partial E/\partial J$, and is
shown to be equal to the proper angular velocity of the black hole
with respect to the Eulerian observers who are at rest in the
stationary time slices.

Our results show that there is a unique black hole solution for each
temperature~$T$ at the boundary~$B$. Assuming the temperature is
positive, then the heat capacity is positive and the black hole
is thermodynamically stable. Unlike the case in (3+1) dimensions,
there is no negative heat capacity instanton, and therefore no
obvious mechanism to allow for the nucleation of black holes
from \adS~space. These conclusions are qualitatively unchanged in
the limit in which~$B\to\infty$ and~$T\to 0$ in such a way
that the black hole size remains fixed.

For both the (3+1) dimensional \SadS~black hole and the (2+1) dimensional
black hole, we derive the temperature by identifying~$E$ with the internal
energy and {\sl assuming} the appropriate expression for entropy as a
function of horizon size. In order
to actually derive the entropy, it is necessary to perform a quantum
(or at least semi--classical) calculation. One approach, which
is based on path integral techniques~[9,18], is to identify the entropy
in the classical approximation with the `microcanonical action' for the
Euclideanized (or complexified) black hole spacetime~[8]. Even without the
assumption of the appropriate form for the entropy, the calculations in
this paper can be interpreted in terms of black hole mechanics.


\section review

We consider a spacetime manifold~$\cal M$ of dimension~$D$
which is topologically the
product of a spacelike hypersurface and a real line
interval,~$\Sigma\times I$. The boundary of~$\Sigma$ is
denoted~$\partial\Sigma = B$. The spacetime metric is~$g_{\mu\nu}$ with
associated curvature tensor~${\cal R}_{\mu\nu\rho\sigma}$ and derivative
operator~$\nabla_{\!\mu}$. The boundary of~$\cal M$, $\partial\cal M$,
consists of initial and final spacelike hypersurfaces $t'$ and~$t''$,
respectively, and a timelike hypersurface~$\B3 = B\times I$ joining
these. The induced metric on the spacelike hypersurfaces $t'$ and~$t''$
is denoted by~$h_{ij}$, and the induced metric on~$\B3$ is denoted
by~$\gamma_{ij}$.\footnote*{We use latin letters $i$, $j$, $k$, $\ldots$ as
indices both for tensors on~$\B3$ and for tensors on a generic
hypersurface~$\Sigma$. The two uses of such indices can be distinguished
by the context in which they occur.}

The gravitational action appropriate for fixation of the
metric on the boundary~$\partial\cal M$ is~[19]
$$S^1 = {1\over2\kappa}\int_{\cal M}\dDx\sqrt{-g}(R-2\Lambda)
  +{1\over\kappa}\int_{t'}^{t''}\dDmx\sqrt{h}\,K
  -{1\over\kappa}\int_\B3 \dDmx \sqrt{-\gamma}\,\Theta
  \ .\eqno\eq$$
Here,~$\kappa$ is a coupling constant and $\Lambda$~is an optional
cosmological constant. For simplicity, we have omitted matter
contributions to the action. The symbol~$\int_{t'}^{t''} d^{D-1}x$ denotes
an integral over the boundary element~$t''$ minus an integral over the
boundary element~$t'$. The function~$K$ is the trace of the extrinsic
curvature~$K_{ij}$ for the boundary elements $t'$ and~$t''$, defined
with respect to the future pointing unit normal. Likewise, $\Theta$~is
the trace of the extrinsic curvature~$\Theta_{ij}$ of the boundary
element~$\B3$, defined with respect to the outward pointing unit normal.

Under variations of the metric, the action~(2.1)
varies according to
$$\eqalignno{\delta S^1&=\hbox{(terms that vanish when the equations of
motion hold)}\cr &\qquad+\int_{t'}^{t''} \dDmx\,P^{ij}\delta h_{ij} +
\int_\B3 \dDmx\,\pi^{ij}\delta \gamma_{ij} \ .&\eq\cr}$$
The coefficient of~$\delta h_{ij}$ in the boundary terms at $t'$ and~$t''$
is, by definition, the gravitational momentum
$$P^{ij} = {1\over2\kappa} \sqrt{h} \bigl( K h^{ij} - K^{ij} \bigr)
\ .\eqno\eq$$
Likewise, the coefficient of~$\delta\gamma_{ij}$ in the boundary term
at~$\B3$ is
$$\pi^{ij} = -{1\over2\kappa} \sqrt{-\gamma} \bigl( \Theta \gamma^{ij}
- \Theta^{ij} \bigr) \ .\eqno\eq$$
In addition to the terms displayed in Eq.~(2.2), $\delta S^1$ includes
an integral over the corner~$t''\cap\B3$ (and an integral over~$t'\cap\B3$)
whose integrand is proportional to the variation of the
`angle'~${\mib u}\cdot{\mib n}$ between the unit normal~${\mib u}$
to~$t''$ (or $t'$) and the unit normal~${\mib n}$ to~$\B3$~[20,21]. We will
not need these terms in the analysis that follows.

The action~$S^1$ yields the classical equations of motion when the
induced metric on~$\partial{\cal M}$ is held fixed in the variational
principle. In general, the functional~$S=S^1-S^0$, where $S^0$~is a
functional of the metric on~$\partial{\cal M}$, also yields the
classical equations of motion when the metric is fixed
on~$\partial{\cal M}$, since in that case $\delta S^0$ vanishes.
For simplicity, we define~$S^0$ to be a functional of~$\gamma_{ij}$
only. The variation~$\delta S$ differs from~$\delta S^1$ of Eq.~(2.2)
only in that $\pi^{ij}$~is replaced
by~$\pi^{ij} - (\delta S^0/\delta\gamma_{ij})$.

Now foliate the boundary element~$\B3$ into ($D-2$)--dimensional
hypersurfaces~$B$ with induced ($D-2$)--metrics~$\sigma_{ab}$. The
($D-1$)--metric~$\gamma_{ij}$ can be written according to the familiar
ADM decomposition~[14] as
$$\gamma_{ij}\,dx^idx^j=-N^2dt^2+\sigma_{ab}
(dx^a+V^a dt)(dx^b+V^b dt) \ ,\eqno\eq$$
where $N$ is the lapse function and $V^a$~is the shift vector. The
corresponding variation of~$\gamma_{ij}$ is~[7]
$$\delta\gamma_{ij} = (-{2} u_iu_j /N) \delta N + (-{2} \sigma_{a\(i}
    u_{j\)} /N)\delta V^a + (\sigma_{\(i}^{a} \sigma_{j\)}^{b})
    \delta\sigma_{ab} \ , \eqno\eq$$
where $u_i$~is the unit normal to the slices~$B$ and
$\sigma^i_a = \delta^i_a$~projects covariant tensors from~$\B3$ to
the slices~$B$. With this result, the
contribution to the variation of~$S$ from the boundary element~$\B3$
becomes
$$\eqalignno{ \delta S\bigr|_{\B3} &= \int_\B3 \dDmx \bigl(\pi^{ij} -
   (\delta S^0/\delta\gamma_{ij}) \bigr) \delta \gamma_{ij} \cr
    &= \int_\B3 \dDmx \sqrt{\sigma}\Bigl( -\varepsilon\delta N +
    j_a\delta V^a + (N/2)s^{ab}\delta\sigma_{ab} \Bigr) \ ,&\eq\cr}$$
where the coefficients of the varied fields are defined by
$$\eqalignno{ \varepsilon &= {2\over N\sqrt{\sigma}} u_i\pi^{ij} u_j +
    {1\over\sqrt{\sigma}} {\delta S^0\over\delta N} \ ,&\eq\cr
    j_a &= -{2\over N\sqrt{\sigma}} \sigma_{ai} \pi^{ij} u_j -
    {1\over\sqrt{\sigma}} {\delta S^0\over\delta V^a} \ ,&\eq\cr
    s^{ab} &= {2\over N\sqrt{\sigma}} \sigma_{i}^a \pi^{ij} \sigma_j^b
    - {2\over N\sqrt{\sigma}} {\delta S^0\over\delta\sigma_{ab}}
    \ .&\eq\cr}$$
The leading terms in Eqs.~(2.8)--(2.10) can be rewritten in terms of the
extrinsic curvature~$k_{ab}$ that is defined by parallel transporting the
unit normal~${\mib n}$ to~$\B3$ across a ($D-2$)--dimensional slice~$B$.
Thus, $k_{ab}$~is the extrinsic curvature of~$B$ considered
as the boundary~$B=\partial\Sigma$ of a spacelike hypersurface~$\Sigma$
whose unit normal~${\mib u}$ is orthogonal to~${\mib n}$. Also let
$P^{ij}$ denote the gravitational momentum for the hypersurfaces~$\Sigma$
that are `orthogonal' to~$\B3$, and let $a_\mu=
u^\nu\nabla_\nu u_\mu$ denote the acceleration of the unit normal~$u_\mu$
for this family of hypersurfaces. The resulting expressions
are~[7]
$$\eqalignno{ \varepsilon &= {1\over\kappa} k - \varepsilon_{\sss0}
    \ ,&\eq\cr
    j_i &= {-2\over\sqrt{h}}\sigma_{ij} P^{jk} n_k - (j_{\sss0})_i
    \ ,&\eq\cr
    s^{ab} &= {1\over\kappa}\bigl( k^{ab} + (n_\mu a^\mu - k)\sigma^{ab}
    \bigr)  - (s_{\sss0})^{ab} \ .&\eq\cr}$$
In these equations, we have expressed the terms
proportional to the functional derivatives of~$S^0$ as $\varepsilon_\sss0$,
$(j_\sss0)_i$, and~$(s_\sss0)^{ab}$. Also note that the indices in
Eq.~(2.12) refer to the hypersurface~$\Sigma$. Thus, $j_i = j_a\sigma^a_i$
where $\sigma^i_a = \delta^i_a$~projects tensors from~$\Sigma$ to~$B$, and
$\sigma_{ij} = \sigma^a_i\sigma_{aj}$.

We will assume that $S^0$~is a
linear functional of the lapse~$N$ and shift~$V^a$, so that
$\varepsilon_\sss0$ and~$(j_\sss0)_i$
are functionals of the two--metric~$\sigma_{ab}$ only~[7]. This condition
implies that~$S^0$ is functionally homogeneous of degree 1 in the lapse
and shift:
$$S^0 = \int_\B3 \dDmx \biggl( N {\delta S^0\over\delta N} + V^a
    {\delta S^0\over\delta V^a} \biggr) \ .\eqno\eq$$
By varying this expression with respect to $N$, $V^a$, and~$\sigma_{ab}$,
we find
$$\int_\B3 \dDmx {N\sqrt{\sigma}\over 2} (s_{\sss0})^{ab} \delta\sigma_{ab}
   = \int_\B3 \dDmx \Bigl( -N\delta \bigl( \sqrt{\sigma}{\varepsilon_\sss0}
   \bigr) + V^a\delta \bigl( \sqrt{\sigma} {(j_\sss0)_a} \bigr) \Bigr)
   \ .\eqno\eq$$
This relationship is useful for the determination of~$(s_{\sss0})^{ab}$
when $\varepsilon_\sss0$ and~$(j_\sss0)_a$ are directly given as functions
of~$\sigma_{ab}$ and its derivatives. For later use, we note that the
variations in~$\sigma_{ab}$ can be split into variations in the
determinant and variations that preserve the determinant
$$\delta \sigma_{ab} = {2\over d}
   \biggl({\sigma_{ab}\over\sqrt{\sigma}}\biggr) \delta\sqrt{\sigma}
   + (\sqrt{\sigma})^{2/d} \delta \biggl( {\sigma_{ab}
   \over (\sqrt{\sigma})^{2/d} }\biggr) \ ,\eqno\eq$$
where $d=D-2$~is the dimension of $B$.

{}From its definition through Eq.~(2.7), $-\sqrt{\sigma}\varepsilon$ is
equal to the time rate of change of the action, where changes in time are
controlled by the lapse function~$N$ on~$\B3$. Thus,
$\varepsilon$~is identified as an energy surface density for the system
and the total quasilocal energy is defined by integration over a
($D-2$)--surface~$B$~[7]:
$$E = \int_B \dDmmx \sqrt{\sigma}\varepsilon  \ .\eqno\eq$$
We also refer to $j_i$ as the momentum surface density and $s^{ab}$ as
the spatial stress~[7].

When there is a Killing vector field~$\mib\xi$ on the boundary~$\B3$,
an associated conserved charge is defined by~[7]
$$Q_\xi=\int_{B } \dDmmx \sqrt{\sigma} (\varepsilon u^i+j^i) \xi_i
   \ .\eqno\eq$$
If there is no matter stress--energy in the neighbourhood
of~$\B3$, $Q_\xi$~is conserved in the sense that $Q_\xi$ is
independent of the particular surface~$B$ (within~$\B3$) that is
chosen for its evalutation~[7]. This property is not shared by the
energy~$E$. If the system contains a rotational symmetry given by a
Killing vector field~${\mib\zeta}$ on~$\B3$, the conserved charge is the
angular momentum~$J = Q_\zeta$. If the ($D-2$)--surface~$B$
is chosen to contain the orbits of~${\mib\zeta}$, then the angular momentum
can be expressed as
$$J = \int_B \dDmmx \sqrt{\sigma} j_i \zeta^i \ .\eqno\eq$$
This is the integral over~$B$ of the $\phi$ component of the momentum
surface density~$j_i$, where $\zeta^i = (\partial/\partial\phi)^i$.

If the Killing vector field~${\mib\xi}$ is timelike, then the negative of
the corresponding charge~(2.18) defines a conserved mass for the system,
$M = - Q_\xi$. If the Killing vector field is also surface forming, then
the mass can be evaluated on a surface~$B$ whose unit normal is
proportional to~${\mib\xi}$. In this case the conserved mass is given by
$$M = \int_B \dDmmx \sqrt{\sigma} N\varepsilon \ ,\eqno\eq$$
where $N$ is the lapse function defined by~${\mib\xi} = N {\mib u}$.
If, in addition, ${\mib\xi}$~restricted to~$B$ has unit norm,
$N=1$, then the conserved mass~$M$ coincides with the
energy~(2.17) of the hypersurface~$\Sigma$ whose boundary is~$B$.
However, if~${\mib\xi}$ does not have unit norm at~$B$, then the mass~$M$
will differ from the energy~$E$. Moreover, the energy~$E$ evaluated
on other slices of~$\B3$ will
not, in general, equal the conserved mass~$M$. These distinctions
between mass and energy are especially important for spacetimes that are
asymptotically \adS, since in that case the magnitude of
the timelike Killing vector field diverges as it approaches infinity.
Thus, the timelike Killing vector does not approach the unit normal to
the (asymptotically) stationary time slices at spatial infinity,
and the mass~$M$ and energy~$E$ do not coincide.

For asymptotically flat or asymptotically \adS~spacetimes,
the ADM charges at infinity, as defined by an analysis of the surface
terms in the Hamiltonian~[14,15,22,23], coincide with (the negative of)
the conserved charges~(2.18). In order to verify this connection,
note that the Hamiltonian derived from the action~(2.1) is~[7]
$$H=\int_\Sigma \dDmx (N{\cal H}+V^i{\cal H}_i) + \int_{B} \dDmmx
   \sqrt{\sigma}(N\varepsilon-V^i j_i) \ .\eqno\eq$$
When the constraints~${\cal H} = 0 = {\cal H}_i$ hold, the first
integral in the Hamiltonian vanishes. Thus the energy~(2.17) is the
value of the Hamiltonian with $N=1$ and~$V^i = 0$ on the boundary~$B$;
that is, the value of the Hamiltonian that generates a unit time
translation on the boundary in the direction orthogonal to the
hypersurface~$\Sigma$. Now push the boundary~$B$ to infinity along an
asymptotically stationary time slice~$\Sigma$. The ADM charges are
defined by the value of the Hamiltonian that generates an evolution
that asymptotically coincides with an asymptotic Killing vector~${\mib\xi}$.
Therefore each ADM charge is given by the boundary
integral in the Hamiltonian~(2.21), where $N$ and~$V^i$ are chosen
such that (asymptotically) $N {\mib u} + V^i(\partial/\partial x^i)
= {\mib\xi}$. On the other hand, setting~${\mib\xi}$ equal to
$N {\mib u} + V^i(\partial/\partial x^i)$ in Eq.~(2.18), we see
that the conserved charge associated with the asymptotic Killing
vector~${\mib\xi}$ is
$$Q_\xi = - \int_B \dDmmx \sqrt{\sigma} (N\varepsilon - V^i j_i )
   \ .\eqno\eq$$
This is the negative of the boundary term in the Hamiltonian~$H$.
Therefore, if~$B$ is taken to infinity, $Q_\xi$~is the negative of
the ADM charge associated with~${\mib\xi}$.
Specifically, the ADM mass and angular momentum agree with the mass~$M$
and angular momentum~$J$ of Eqs.~(2.20) and~(2.19), in the limit that~$B$
is taken to infinity.

It should be recognized that both the conserved charges~(2.18) and the
ADM charges depend on the normalization of the (asymptotic) Killing
vector
field. Thus, the charge associated with the Killing vector~$c{\mib\xi}$,
where $c$~is a constant, is equal to the product of~$c$ and the charge
associated with~${\mib\xi}$. A second and more subtle ambiguity in the
charges arises
because of the presence of the terms $\varepsilon_\sss0$ and~$(j_\sss0)_i$
in the energy surface density and momentum surface density. These terms
depend on the boundary metric~$\sigma_{ab}$. In the
standard ADM analysis at infinity, these terms are effectively chosen
such that the mass and angular momentum vanish for a flat time slice
of flat Minkowski spacetime if~$\Lambda = 0$, or for a static time slice
of \adS~spacetime if~$\Lambda<0$. (Alternatively, in (2+1)
dimensions with~$\Lambda<0$, mass and angular momentum are conveniently
chosen to vanish for a static time slice of the zero mass black hole
solution.) In the asymptotically flat case, this choice of a zero
point configuration  for the mass and angular momentum is typically built
into the analysis through the use of coordinate derivatives acting on the
metric tensor components, where the coordinate system is asymptotically
Cartesian.


\section \SadS\ spacetimes

In this section, we will consider the \SadS~black hole solutions to
(3+1)~dimensional Einstein gravity with a negative cosmological
constant~$\Lambda$. We will adopt units for which~$\kappa=8\pi$
(thus~$G=1$), and also set $\Lambda = -3/\ell^2$. The metric written in
static spherical coordinates is
$$ds^2 = -N^2(r) \,dt^2 + f^{-2}(r) dr^2 +
      r^2(d\theta^2+\sin^2\!\theta\,d\phi^2) \ ,\eqno\eq$$
where
$$ N^2(r) = f^{2}(r) = {\textstyle\bigl(1-2m/r + r^2/\ell^2 \bigr)}
     \ .\eqno\eq$$
Let~$\Sigma$ be the interior of a $t={\rm const}$ slice with
two--boundary~$B$ specified by~$r=R={\rm const}$. The
term~$\varepsilon_\sss0$
in the energy surface density is a function of~$R$, and the
term~$(j_\sss0)_a$
in the momentum surface density is chosen to be zero.
If variations in the metric~$\sigma_{ab}$ are restricted to variations
in the radius~$R$, then Eqs.~(2.15) and~(2.16) imply
$$\int_\B3 d^3x {N\over2} \sigma_{ab}(s_\sss0)^{ab} \delta\sqrt{\sigma}
   = - \int_\B3
   d^3x N {\partial(\sqrt{\sigma}\varepsilon_\sss0)\over
   \partial\sqrt{\sigma}} \delta\sqrt{\sigma} \ .\eqno\eq$$
It follows that
$$\sigma_{ab}(s_\sss0)^{ab} = - 2{\partial(R^2\varepsilon_\sss0) \over
   \partial (R^2) } \ ,\eqno\eq$$
since~$N$ is constant on~$\B3$.

A straightforward calculation of the
trace of the extrinsic curvature~$k_{ab}$ for the spherical boundary~$r=R$
in \SadS~spacetime yields
$$k = -{2f(R)\over R} =
 -{2\over R} \sqrt{1 - 2m/R + R^2/\ell^2} \ .\eqno\eq$$
The acceleration of the unit normal~$u_\mu$ satisfies
$$n_\mu a^\mu = {f(R)\,N^\prime(R)\over N(R)} =
 {m/R + R^2/\ell^2 \over R\sqrt{1 - 2m/R + R^2/\ell^2} } \ ,\eqno\eq$$
where the prime indicates a radial derivative. From these results,
the energy surface density~(2.11) is given by
$$\varepsilon = -{1\over4\pi R} \sqrt{1 - 2m/R + R^2/\ell^2} -
   \varepsilon_\sss0(R) \ ,\eqno\eq$$
and the total energy~(2.17) is
$$E = -R\sqrt{1 - 2m/R + R^2/\ell^2} - 4\pi R^2\varepsilon_\sss0(R)
   \ .\eqno\eq$$
Also, the trace of the spatial stress~(2.13) is given by
$$ \sigma_{ab}s^{ab} = {1\over4\pi R} \biggl( {1 - m/R + 2R^2/\ell^2 \over
   \sqrt{1 - 2m/R + R^2/\ell^2} }\biggr) + {1\over R}
   {\partial(R^2\varepsilon_\sss0) \over\partial(R) } \ ,\eqno\eq$$
where Eq.~(3.4) has been used. For the boundary~$r=R$ of the
$t={\rm const}$ slices of \SadS, the momentum surface density~(2.12)
vanishes.

If we choose
$$\varepsilon_\sss0(R)  = -{1\over4\pi R} \sqrt{1 + R^2/\ell^2}
\ ,\eqno\eq$$
then the energy surface density~$\varepsilon$ and the energy~$E$ vanish for
\adS~spacetime ($m=0$, $\ell$~finite). Also note that with this choice, as
$R\to\infty$ the energy vanishes: $E\sim m\ell/R \to 0$.
Another natural choice is $\varepsilon_\sss0 = -1/(4\pi R)$. In that case
$\varepsilon$ and~$E$ vanish for flat spacetime ($m=0$, $\ell\to\infty$),
and $E\sim -R^2/\ell \to \infty$ as~$R\to\infty$. The simplest and most
convenient choice for~$\varepsilon_\sss0$ is~$\varepsilon_\sss0 = 0$. We
will leave~$\varepsilon_\sss0(R)$ unspecified in the analysis that follows.

The entropy~${\cal S}$ of any stationary black hole in (3+1) dimensional
Einstein gravity is one-quarter the area of its event horizon. (This includes
black holes that are distorted by stationary matter fields relative to
the standard Kerr or Kerr--\adS~family.) This conclusion was first
reached by Bekenstein~[24] apart from an overall numerical factor, and
is derived as a general result in Ref.~[8]. Moreover, it is now well
recognized~[25] that black hole entropy depends only on the geometry of
the horizon (in the classical approximation). This suggests that the
expression for black hole entropy is independent of the asymptotic
behavior of the gravitational field, or the presence of external matter
fields. Thus, for the \SadS~black hole~(3.1), the entropy is ${\cal S}
= \pi (r_\sssH)^2$, where~$r_\sssH$ satisfies
$$N^2(r_\sssH)=1 - 2m/r_\sssH + (r_\sssH)^2/\ell^2 = 0 \ . \eqno\eq$$
For a given~$m$, there is a unique real solution for the event
horizon~$r_\sssH$. The expressions~(3.5)--(3.9) are real and physically
meaningful only for~$r_\sssH < R$.

Now identify the energy~(3.8) as the thermodynamic internal energy for the
\SadS~black hole spacetime within the boundary~$R$; view~$E$ as a
function of the entropy ${\cal S} = \pi (r_\sssH)^2$ and the boundary
area~$4\pi R^2$. The corresponding temperature is
$$ T \equiv \biggl( {\partial E\over \partial{\cal S}}   \biggr)
       = {1\over\sqrt{1- 2m/R + R^2/\ell^2}} \biggl( {1 + 3(r_\sssH)^2/\ell^2
       \over 4\pi r_\sssH}\biggr) \ . \eqno\eq$$
The second factor in this expression is just~$1/(2\pi)$ times the surface
gravity~$\kappa_\sssH$ of the black hole, where
$$\kappa_\sssH^2 = -{1\over2}(\nabla^\mu\chi^\nu)(\nabla_\mu\chi_\nu) =
    (\partial_i N)h^{ij}(\partial_j N)
    \qquad\hbox{(evaluated on the horizon)} \ ;\eqno\eq$$
with~$\chi$ being a Killing vector normal to the horizon.
The first factor in Eq.~(3.12) is the inverse of the lapse function
$1/N = \sqrt{-g^{tt}}$ evaluated at~$R$, and is the
Tolman redshift factor for temperature in a stationary gravitational
field~[13].
Therefore, the temperature at~$R$ is the product of $\kappa_\sssH/(2\pi)$
and the redshift factor:
$$T(R) = {1\over 2\pi} {\kappa_\sssH \over N(R)} \ .\eqno\eq$$
For a given size black hole, the temperature~$T$ redshifts to zero as
$R\to\infty$ (assuming the cosmological constant is nonzero). Note that
although the surface gravity~(3.13) depends on
the scale of the coordinate~$t$ that labels the stationary time slices,
the temperature~(3.14) does not. Also observe that the temperature is
independent of the choice of function~$\varepsilon_\sss0$.

{}From the energy~$E$, we can also define a thermodynamic surface pressure by
$$ \Pres \equiv -\biggl( {\partial E\over\partial(4\pi R^2)} \biggr)
        =  {1\over8\pi R} \biggl( {1 - m/R + 2R^2/\ell^2 \over
   \sqrt{1 - 2m/R + R^2/\ell^2} }\biggr) + {\partial(R^2\varepsilon_\sss0)
   \over\partial(R^2) } \ . \eqno\eq$$
This is precisely one half of the trace of the spatial stress tensor~(3.9):
$$\Pres = {1\over 2}\sigma_{ab} s^{ab} \ .\eqno\eq$$
The surface pressure does depend on the function~$\varepsilon_\sss0$. The
definitions for temperature and surface pressure are captured in the
first law of thermodynamics, namely,
$dE = T d{\cal S} - \Pres d(4\pi R^2)$.

The heat capacity at constant surface area~$4\pi R^2$ is defined
by
$$C_R \equiv \biggl({\partial E\over\partial T}\biggr) \ ,\eqno\eq$$
where the energy~$E$ is expressed as a function of $T$ and~$R$.
The energy~(3.8) and temperature~(3.12) can be expressed
as functions of $r_\sssH$ and~$R$ by eliminating~$m$ through Eq.~(3.11).
Then the heat capacity can be written as
$$C_R = \biggl({\partial E\over\partial r_\sssH}\biggr) \biggl({\partial T
   \over\partial r_\sssH}\biggr)^{-1} \ .\eqno\eq$$
It is
straightforward to show that~$E$ is a monotonically increasing function
of~$r_\sssH$ for~$0<r_\sssH<R$, so that ${\partial E/\partial r_\sssH}$
is strictly positive. On the other hand, the temperature~$T$ is a
positive function of~$r_\sssH$ with $T\to\infty$ both as $r_\sssH\to 0$
and as~$r_\sssH\to R$. It can be shown that in the range $0<r_\sssH<R$,
$T$~has a single extremum\footnote*{ $\partial T/\partial r_\sssH$
is equal to a positive function of~$r_\sssH$ times a fifth order polynomial
in~$r_\sssH$. This  polynomial has precisely one positive root,
because the constant term is negative and the signs of the coefficients
of all higher powers of~$r_\sssH$ are positive. The single root lies in
the range $0<r_\sssH<R$ since the polynomial is positive at~$r_\sssH = R$.}
which is a minimum~$T_0$. Therefore,
Eq.~(3.12) has no solution for~$r_\sssH$ when~$T<T_0$, and has two
solutions for~$r_\sssH$ when~$T>T_0$. Physically, this means that
there are no \SadS~black hole solutions with temperature at~$r=R$ less
than~$T_0$. If the temperature at~$r=R$ is fixed to a value less than~$T_0$
then the system will be dominated by thermal radiation in an
\adS~background. If the temperature at~$r=R$ is fixed to a value greater
than~$T_0$, there are two black hole solutions, a small black hole
with $\partial T/\partial r_\sssH <0$ and a large black hole with
$\partial T/\partial r_\sssH>0$. Since the sign of the heat capacity~(3.18)
coincides with the sign of $\partial T/\partial r_\sssH$, only
the larger of the two black holes is thermodynamically stable.

The preceeding analysis is qualitatively unchanged in the limit of a
vanishing cosmological constant, $\Lambda=0$ ($\ell\to\infty$).
Previous work~[1,16] on the $\Lambda=0$ case shows that the Euclidean
section of the small black hole
with $C_R<0$ is an instanton that dominates the semiclassical evaluation
of the rate of nucleation of black holes in a cavity of size~$R$ and
temperature~$T$. The large black hole with $C_R>0$ is the end result
of the nucleation process.

Consider the heat capacity with the limit $R\to\infty$ taken in such a
way that the black hole size~$r_\sssH$ remains fixed. If the cosmological
constant is negative ($\ell$ is finite), then the temperature~$T$ will
go to zero in this limit; if the cosmological constant is zero
($\ell\to\infty$), then the temperature will go to $1/(4\pi r_\sssH)$.
In general, the result for the heat capacity is
$$\lim_{R\to\infty} C_R = -2\pi r_\sssH^2 \biggl(
   {1+3r_\sssH^2/\ell^2 \over 1-3r_\sssH^2/\ell^2} \biggr) \ .\eqno\eq$$
Now, for large~$R$, the minimum~$T_0$ of the function~$T(r_\sssH)$
occurs at $r_\sssH = \ell/\sqrt{3}$. If $\ell$ is finite, then
Eq.~(3.19) confirms that for the large black hole ($r_\sssH>\ell/\sqrt{3}$)
the heat capacity is positive, while for the small black hole
($r_\sssH<\ell/\sqrt{3}$) the heat capacity is negative. However, if
the cosmological constant vanishes ($\ell\to\infty$), then the minimum~$T_0$
becomes infinite. In this case the horizon size for the large black
hole must become infinite as $R\to\infty$. Therefore it is not possible
to take the limit $R\to\infty$ while keeping the large black hole size
fixed. When the cosmological constant vanishes, only the small black
hole instanton can survive the $R\to\infty$ limit. Correspondingly, for
$\ell\to\infty$ the heat capacity~(3.19) is strictly negative.

The above results indicate that a black hole in infinite space
($R\to\infty$) can be in thermal equilibrium if the cosmological
constant is negative, but not if the cosmological constant is zero.
The same conclusion has been reached by Hawking and Page~[11].

Finally, consider the conserved charges associated with the
\SadS\ black hole. Since the momentum surface density vanishes,
there is no angular momentum~(2.19) as expected. The conserved mass~(2.20)
differs from the energy~$E$ by a factor of the lapse function
at~$r=R$:
$$\eqalignno{ M &= N(R)\,E \cr
    &= -R\bigl( 1 - 2m/R + R^2/\ell^2 \bigr) - 4\pi R^2
    \sqrt{ 1 - 2m/R + R^2/\ell^2 } \varepsilon_\sss0(R) \ .&\eq\cr}$$
We can view~$M$ as a function of the entropy ${\cal S}=\pi (r_\sssH)^2$
and the boundary size~$R$ by expressing~$m$ in terms of~$r_\sssH$
through Eq.~(3.11). Then the mass varies with entropy
according to the relationship
$$ {\partial M\over\partial{\cal S}} = N{\partial E
     \over\partial{\cal S}} + E {\partial N\over\partial{\cal S}}
     = {\kappa_\sssH \over 2\pi} + {E\over 2\pi r_\sssH}
     {\partial N\over \partial r_\sssH} \ ,\eqno\eq$$
where the result (3.12), (3.14) has been used. Note that
$\partial N/\partial r_\sssH = -2\pi T r_\sssH/R$. If
$E(\partial N/\partial r_\sssH)$ vanishes as $R\to\infty$, then in
this limit $\partial M/\partial{\cal S}$ is the product of $1/(2\pi)$ and
the surface gravity~$\kappa_\sssH$. This is indeed the case if we
choose~$\varepsilon_\sss0$ as in Eq.~(3.10), since then $E\to 0$ as
$R\to\infty$. Moreover, with this choice for~$\varepsilon_\sss0$, the
conserved mass reduces to $M=m$ in the $R\to\infty$ limit. However, it
is not correct to interpret $M=m$ as the thermodynamic internal energy
and $\kappa_\sssH/(2\pi)$ as the temperature at infinity---the physical
temperature~(3.14) redshifts to zero at infinity. In fact,
$\kappa_\sssH/(2\pi)$ is the temperature at the spatial location where
the lapse function~$N(R)$ (the inverse of the redshift factor) equals
unity. Recall that the conserved mass~(2.20) and the surface gravity~(3.13)
depend on the choice of scale for the time coordinate~$t$.
Likewise, the location at which $N(R) = 1$ depends on the choice
of time coordinate.

The heat capacity (3.17) can be expressed as
$$C_R = \biggl({\partial E\over\partial T}\biggr) =
   \biggl({\partial(NE)\over\partial r_\sssH}
   - E {\partial N\over\partial r_\sssH}\biggr)
   \biggl({\partial(NT)\over\partial r_\sssH}
   - T {\partial N\over\partial r_\sssH}\biggr)^{-1} \ .\eqno\eq$$
If $E(\partial N/\partial r_\sssH)$ and $T(\partial N/\partial r_\sssH)$
vanish as $R\to\infty$, then in this limit the heat capacity becomes
$$\lim_{R\to\infty} C_R = \lim_{R\to\infty}
   \biggl({\partial M\over\partial r_\sssH}\biggr)
   \biggl({\partial (\kappa_\sssH/2\pi)\over \partial r_\sssH}\biggr)^{-1}
   =  \biggl({\partial m\over\partial(\kappa_\sssH/2\pi)}\biggr)
   \ .\eqno\eq$$
Again, the term $E(\partial N/\partial r_\sssH)$ can be dropped
if~$\varepsilon_\sss0$ is chosen as in Eq.~(3.10).  Whether or not the
term $T(\partial N/\partial r_\sssH) = -2\pi T^2 r_\sssH/R$ can
be dropped as $R\to\infty$ depends on how the limit is taken.
If $r_\sssH$ is held fixed in the limit $R\to\infty$, then
$T(\partial N/\partial r_\sssH)$ indeed vanishes (for both~$\ell$
finite and $\ell\to\infty$). Thus, with this definition for the
limit, the expression $\partial m/\partial(\kappa_\sssH/2\pi)$
correctly yields the heat capacity~(3.19). On the other hand, if
the temperature~$T$ is held fixed as $R\to\infty$, then
$T(\partial N/\partial r_\sssH)$ vanishes only for the small black
hole. For the large black hole, $T(\partial N/\partial r_\sssH)$
does not vanish because $r_\sssH\to\infty$ as $R\to\infty$.


\section (2+1) dimensional black hole

We now consider (2+1) dimensional Einstein gravity with a
negative cosmological constant~$\Lambda$.
We will adopt units in which $\kappa = \pi$ and
set $\Lambda = -1/\ell^2$. The axially symmetric black hole solution
obtained by Ba\~nados {\sl et al.}~[12,17] written in stationary coordinates
is
$$ds^2=-N^2(r)\,dt^2+f^{-2}(r)\,dr^2+r^2\bigl(V^\phi(r)\,dt+
    d\phi\bigl)^2 \ ,\eqno\eq$$
where
$$N^2(r)=f^2(r)=-m+\Bigl({r\over \ell}\Bigr)^2+\Bigl({j\over2r}\Bigr)^2
   \qquad{\rm and}\qquad V^\phi(r)=-{j\over 2r^2} \ .\eqno\eq$$
An analysis of the Hamiltonian for (2+1) gravity shows that $m$ and~$j$
are the ADM mass and angular momentum at infinity~[17,23]. The mass
parameter~$m$ also can be expressed in terms of the initial energy
density of a disk of collapsing dust~[26] in \adS~space or alternatively
in terms of Casimir invariants in a gauge--theoretic formulation of
(2+1)--dimensional general relativity~[27]. As with the Kerr solution,
the lapse function~$N(r)$ for the (2+1) black hole vanishes for two
values of~$r$, namely $r_+$ and~$r_-$, where
$$(r_\pm)^2 = {m\ell^2\over2} \pm {\ell\over2}\sqrt{m^2\ell^2 - j^2}
     \ .\eqno\eq$$
The larger of these,~$r_+$, is specified as the black hole horizon. Such
a horizon exists only for $m>0$ and~$|j|\le m\ell$.
(When~$|j|=m\ell$, $r_+=r_-$.)

Let~$\Sigma$ be the interior of a~$t={\rm const}$ slice with boundary~$B$
specified by~$r=R={\rm const}$. The term~$\varepsilon_\sss0$ in the
energy surface density is a function of~$R$, and the term~$(j_\sss0)_i$
in the momentum surface density is chosen to be zero. Eqs.~(2.15)
and~(2.16) imply
$$\sigma_{ab}(s_\sss0)^{ab} = - {\partial(R\varepsilon_\sss0) \over
   \partial R } \ ,\eqno\eq$$
since variations in the metric~$\sigma_{ab}$ consist of variations
in the surface `area'~$2\pi R$.

Straightforward calculations yield
$$k = -{f(R)\over R} =
-{1\over R}\sqrt{-m + R^2/\ell^2 + j^2/(4R^2)} \eqno\eq$$
for the trace of the extrinsic curvature of~$B$ and, as in equation~(3.6),
$$n_\mu a^\mu = {f(R)\,N^\prime(R)\over N(R)} =
{ R^2/\ell^2 - j^2/(4R^2) \over
R\sqrt{-m + R^2/\ell^2 + j^2/(4R^2)} } \eqno\eq$$
for the acceleration at~$B$ of the unit normal~$u_\mu$ to the stationary
time slices. From these results the energy surface density~(2.11) is
given by
$$\varepsilon = -{1\over\pi R} \sqrt{-m + R^2/\ell^2 + j^2/(4R^2)}
    -\varepsilon_\sss0(R) \ ,\eqno\eq$$
and the total energy~(2.17) is
$$E = -2 \sqrt{-m + R^2/\ell^2 + j^2/(4R^2)} -
    2\pi R\varepsilon_\sss0(R) \ .\eqno\eq$$
The trace of the spatial stress~(2.13) is
$$\sigma_{ab}s^{ab} = {1\over\pi R} \biggl( { R^2/\ell^2 - j^2/(4R^2)
   \over \sqrt{-m + R^2/\ell^2 + j^2/(4R^2)} }\biggr)  +
   {\partial( R\varepsilon_\sss0) \over\partial R} \ ,\eqno\eq$$
where Eq.~(4.4) has been used.

For the (2+1) black hole, the only nonzero component of the
gravitational momentum~(2.3) is
$P^{r\phi} = -rf(r)(V^\phi)^\prime/(4\pi N(r)) = -j/(4\pi r^2)$.
It follows that the momentum surface density~(2.12) is
$$j_\phi = {j\over 2\pi R} \ ,\eqno\eq$$
and the total angular momentum~(2.19) associated with the Killing vector
field~$\partial/\partial\phi$ is equal to the parameter~$j$:
$$J = j \ .\eqno\eq$$
This result is independent of the boundary size~$R$.
This is because the difference in~$J$ between two surfaces $B_1$ and~$B_2$
of some slice~$\Sigma$ is given by the matter momentum density
in the~$\partial/\partial\phi$ direction, integrated over the region
of~$\Sigma$ bounded by $B_1$ and~$B_2$~[7]. Since the matter momentum
density vanishes for the (2+1) black hole, the angular momentum~$J$ is
the same for any surface~$B$ within the stationary slice~$\Sigma$.

The energy~$E$ and angular momentum~$J$ will vanish for the zero mass
black hole (the metric (4.1) with~$m=0$, $j=0$) if we choose
$$\varepsilon_\sss0(R) = -{1\over \pi\ell} \ .\eqno\eq$$
Another natural choice is
$\varepsilon_\sss0 = -\sqrt{1+R^2/\ell^2}/(\pi R)$. In this case,
$E$ and~$J$ vanish for \adS~spacetime.
With either of these choices, we find
$E\sim m\ell/R$ for~$R\gg \ell$ so the energy vanishes as~$R\to\infty$.

The results of Ref.~[12] show that the entropy~${\cal S}$ of the
(2+1) black hole~(4.1) is~$4\pi r_+$: twice the `area' of its event
horizon. General arguments like those used in (3+1) dimensions show
that $4\pi r_+$ is the entropy for any stationary black hole in (2+1)
dimensional Einstein gravity. We will accept this result and identify
the energy~$E$ as the thermodynamic internal energy for the black hole
spacetime within the spatial region bounded by~$r=R$. Then the
corresponding temperature is given by
$$T \equiv \biggl( {\partial E\over\partial{\cal S}} \biggr) =
    {1\over \sqrt{-m + R^2/\ell^2 + (J/2R)^2} }\biggl(
    {(r_+/\ell)^2 - (J/2r_+)^2 \over 2\pi r_+} \biggr) \ .\eqno\eq$$
Here, $E$~is treated as a function of entropy~${\cal S}= 4\pi r_+$,
angular momentum~$J$, and boundary `area'~$2\pi R$ by solving Eq.~(4.3)
for~$m$ as a function of~$r_+$ and setting~$j=J$. The first factor
in this expression for~$T$ is the Tolman redshift factor~$1/N(R)$ for
temperature in a stationary gravitational field. The second factor
is the product of~$1/(2\pi)$ and the surface gravity~(3.13). Thus, just
as in (3+1) dimensions, the temperature at~$R$ is given by
$$T(R) = {1\over2\pi} {\kappa_\sssH\over N(R)} \ .\eqno\eq$$
For a given size black hole, the temperature redshifts to zero
as~$R\to\infty$.

The thermodynamic surface pressure defined by the energy~$E$ is equal to the
trace of the spatial stress~(4.9):
$$\Pres \equiv -\biggl( {\partial E\over\partial(2\pi R)}\biggr) =
   \sigma_{ab} s^{ab} \ .\eqno\eq$$
This result and the result~(3.16) for (3+1) dimensional black holes
differ by a factor of one half. The difference stems from the fact that the
surface pressure is defined by variations in~$E$ with respect to variations
in the boundary metric~$\sigma_{ab}$ that preserve the conformally
invariant part of the
metric~$\sigma_{ab}/(\sqrt{\sigma})^{2/d}$. The factor of~$2/d$ in
Eq.~(2.16) shows that in (2+1) dimensions ($d=1$) the variations
in~$\sqrt{\sigma}$ have an extra factor of two relative
to the variations in~$\sqrt{\sigma}$ for (3+1) dimensions~($d=2$).

Next, we compute the thermodynamic chemical potential conjugate to angular
momentum~$J$. It is defined by
$$\eqalignno{\omega \equiv \biggl( {\partial E\over\partial J} \biggr)
    &= { J/(2r_+^2) - J/(2R^2) \over \sqrt{-m + R^2/\ell^2 + (J/2R)^2} }
    \cr &= { V^\phi(R) - V^\phi(r_+) \over N(R) } \ .&\eq\cr}$$
Observe that the angular velocity of the black hole
horizon with respect to the spatial coordinate system is~$-V^\phi(r_+)$.
This can be verified by showing that the Killing vector field
$\chi^\mu = (\partial/\partial t)^\mu -
V^\phi(r_+)\,(\partial/\partial\phi)^\mu$ is null on the horizon~[28]. By
definition, the shift vector~$V^\phi(R)$ is the angular velocity of the
spatial coordinate system relative to the Eulerian observers at~$R$ whose
four velocities are orthogonal to the stationary time slices~$t = {\rm
const}$. Therefore, $V^\phi(R) - V^\phi(r_+)$~is the angular velocity
of the black hole with respect to the Eulerian observers at~$R$.
This is an improper angular velocity, in the sense that it is taken
with respect to coordinate time~$t$. But coordinate time~$t$ is related
to the proper time of the Eulerian observers at~$r=R$ by a factor of the
lapse function~$N(R)$. Therefore we see that the chemical potential~$\omega$
of Eq.~(4.16) is the proper angular velocity of the  black
hole with respect to the Eulerian observers at the boundary~$B$ of
the system. This is the expected result~[4,8].

The definitions for temperature, surface pressure, and chemical potential
are captured in the first law of thermodynamics for the (2+1) black
hole, namely, $dE = T d{\cal S} + \omega dJ - \Pres d(2\pi R)$.

The heat capacity at constant surface `area'~$2\pi R$ and constant
angular momentum~$J$ is
$$C_{R,J} \equiv \biggl( {\partial E\over\partial T}\biggr) =
    \biggl( {\partial E\over\partial r_+} \biggr)
    \biggl( {\partial T\over\partial r_+} \biggr)^{-1} \ ,\eqno\eq$$
where the energy~(4.8) and temperature~(4.13) are expressed as functions
of $r_+$, $R$, and~$J$. The square root factor that appears in both
$E$ and~$T$ is the lapse function evaluated at~$R$, which can be
expressed in terms of~$r_+$ as
$$N(R) = \bigl[ -(J/2r_+)^2 - (r_+/\ell)^2 + (R/\ell)^2 +
    (J/2R)^2 \bigr]^{1/2} \ .\eqno\eq$$
It is straightforward to show that the derivative~%
$\partial T/\partial r_+$ is positive, so that the temperature
is a monotonically increasing function of~$r_+$. This means that
there is a unique black hole with a given temperature~$T(R)$
and a given angular momentum~$J$. Moreover, note that
for~$R> r_+$, the inequalities
$r_+ > r_+^2/R \geq m\ell^2/(2R) \geq |J|\ell/(2R)$ follow from
the explicit form~(4.3) of~$r_+$ and the condition~$|J|\leq m\ell$.
Thus, we find that~$r_+$ is limited to the range
$$R> r_+ > |J|\ell/(2R) \ .\eqno\eq$$
The lapse function~(4.18) is real for~$r_+$ in this range, and
vanishes at the endpoints $r_+ = R$ and~$r_+ = |J|\ell/(2R)$.
It follows that the temperature~$T$ increases monotonically from~%
$-\infty$ at~$r_+ = |J|\ell/(2R)$ to~$+\infty$ at~$r_+ = R$. The
temperature vanishes for $r_+ = \sqrt{|J|\ell/2}$, so that~$T> 0$
for~$r_+>  \sqrt{|J|\ell/2}$. Now, a simple calculation shows
that  the derivative $\partial E/\partial r_+$ is positive
for $r_+ > \sqrt{|J|\ell/2}$ and negative for
$r_+ < \sqrt{|J|\ell/2}$. Therefore the sign of the heat capacity~(4.17)
is the same as the sign of the temperature. In conclusion,
there is a unique black hole with positive temperature~$T$ at~$r=R$
and angular momentum~$J$, and this black hole is
thermodynamically stable~($C_{R,J}>0$). Unlike the case in (3+1)
dimensions, for~$T>0$ there is no negative heat capacity
instanton.

{}From Eq.~(2.18) we find that the conserved mass associated with the
Killing vector field ${\mib\xi} = (\partial/\partial t)$ is equal to
$$M = -Q_\xi = N(R)\,E + {J^2\over 2R^2} \ .\eqno\eq$$
Note that Eq.~(2.20) cannot be used for this calculation, since~${\mib\xi}$
is not orthogonal to the hypersurface~$\Sigma$. Consider~$M$ to be a function
of the entropy~${\cal S} = 4\pi r_+$, angular
momentum~$J$, and boundary size~$R$. If we choose the function~%
$\varepsilon_\sss0$ as in Eq.~(4.12), then for a given size black
hole~$E$ vanishes as~$R\to\infty$. We then find, as in (3+1)
dimensions, that $M\to m$ as~$R\to\infty$ and
$$\lim_{R\to\infty} \biggl(
   {\partial M\over\partial{\cal S}}\biggr)
   = {\kappa_\sssH\over 2\pi} \ .\eqno\eq$$
The heat capacity can be expressed as in Eq.~(3.22) (with~$r_\sssH$
replaced by~$r_+$) and the term~$E(\partial N/\partial r_+)$ can
be dropped assuming~$\varepsilon_\sss0$ is
chosen appropriately. The term $T(\partial N/\partial r_+) =
-2\pi T^2$ can be dropped if~$T=0$. Thus we obtain
$$ \lim_{ R\to\infty } C_{R,J} =
   {\partial m\over\partial (\kappa_\sssH/2\pi)}
   =4\pi r_+ \biggl(
   {(r_+/\ell)^2 - (J/2r_+)^2 \over (r_+/\ell)^2 + 3(J/2r_+)^2}
   \biggr) \ ,\eqno\eq$$
where $T\to 0$ as~$R\to\infty$ in such a way that the black hole size~$r_+$
remains fixed.
If~$r_+$ satisfies $r_+>\sqrt{|J|\ell/2}$, then Eq.~(4.22)
shows that the heat capacity is positive in the~$T\to 0$,
$R\to\infty$ limit. This agrees with the general result $C_{R,J}>0$
for~$T>0$.


\section acknowledgments

We would like to thank B.~P. Jensen and J.~W. York for helpful remarks.
Also, J.~C. and R.~B.~M. appreciate support from the Natural Sciences and
Engineering Council of Canada (NSERC).


\bigbreak\centerline{REFERENCES}\nobreak\medskip\frenchspacing
\item{[1]} J.~W. York, {\sl Phys. Rev. D. \bf 33} 2092 (1986).
\item{[2]} B.~F. Whiting and J.~W. York, {\sl Phys. Rev. Lett. \bf 61}
1336 (1988).
\item{[3]} J.~D. Brown, G.~L. Comer, E.~A. Martinez, J. Melmed,
B.~F. Whiting, and J.~W. York, {\sl Class. Quantum Grav. \bf 7}
1433 (1990).
\item{[4]} J.~D. Brown, E.~A. Martinez, and J.~W. York, {\sl Phys.
Rev. Lett. \bf 66} 2281 (1991).
\item{[5]} O.~B. Zaslavskii, {\sl Phys. Lett. A. \bf 152} 463 (1991).
\item{[6]} O.~B. Zaslavskii, {\sl Class. Quantum Grav. \bf 8} L103 (1991).
\item{[7]} J.~D. Brown and J.~W. York, {\sl Phys. Rev. D. \bf 47} 1407
(1993).
\item{[8]} J.~D. Brown and J.~W. York, {\sl Phys. Rev. D. \bf 47} 1420
(1993).
\item{[9]} S.~W. Hawking in {\sl General Relativity}, edited by
S.~W. Hawking and W. Israel (Cambridge University Press, Cambridge,
England, 1979).
\item{[10]} R.~M. Wald, in {\sl Black Hole Physics}, edited by
V.~DeSabbata and Z. Zhang (Kluwer Academic Publishers, Dordrecht, 1992).
\item{[11]} S.~W. Hawking and D.~N. Page, {\sl Commun. Math. Phys.
\bf 87} 577 (1983).
\item{[12]} M.~Ba\~nados, C.~Teitelboim and J.~Zanelli, {\sl Phys. Rev.
Lett.\bf 69} 1849 (1992).
\item{[13]} R.~C. Tolman, {\sl Phys. Rev. \bf 35} 904 (1930).
\item{[14]} R.~Arnowitt, S.~Deser, and C.~W. Misner, in {\sl Gravitation:
An Introduction to Current Research}, edited by L.~Witten (Wiley, New
York, 1962).
\item{[15]} T.~Regge and C.~Teitelboim, {\sl Ann. of Phys. \bf 88} 286
(1974).
\item{[16]} D.~J. Gross, M.~J. Perry, and L.~G. Yaffe, {\sl Phys. Rev.
D. \bf 25} 330 (1982).
\item{[17]} M.~Ba\~nados, M.~Henneaux, C.~Teitelboim and J.~Zanelli,
{\sl Phys. Rev. D. \bf 48} (1993) 1506.
\item{[18]} G.~W. Gibbons and S.~W. Hawking, {\sl Phys. Rev. D. \bf 15}
2752 (1977).
\item{[19]} J.~W. York, {\sl Found. Phys. \bf 16} 249 (1986).
\item{[20]} G.~Hayward, {\sl Phys. Rev. D. \bf 47} 3275 (1993).
\item{[21]} J.~D. Brown and J.~W. York, manuscript in preparation.
\item{[22]} M.~Henneaux and C.~Teitelboim, {\sl Commun. Math. Phys.
\bf 98} 391 (1985).
\item{[23]} J.~D. Brown and M.~Henneaux, {\sl Commun. Math. Phys.
\bf 104} 207 (1986).
\item{[24]} J.~D. Bekenstein, {\sl Phys. Rev. D. \bf 7} 2333 (1973).
\item{[25]} D.~Sudarsky and R.~M. Wald, {\sl Phys. Rev. D. \bf 46}
1453 (1992); T.~Jacobson and R.~C. Meyers, {\sl Phys. Rev. Lett.
\bf 70} 3684 (1993); M.~Ba\~nados, C.~Teitelboim, and J.~Zanelli,
{\sl Phys. Rev. D. \bf 49} (1994) 975; R.~M. Wald,
{\sl Phys. Rev. D. \bf 48} 3427 (1993); M. Visser, {\sl Phys. Rev. D.
\bf 48} 5697 (1993); M.~Ba\~nados, C.~Teitelboim, and J.~Zanelli,
{\sl Phys. Rev. Lett. \bf 72} (1994) 957; T.~Jacobson, G.~Kang,
and R.~C. Meyers, ``On black hole entropy".
\item{[26]}S.F. Ross and R.B. Mann, Phys. Rev. {\bf D47} 3319 (1993).
\item{[27]}D. Cangemi, M. Leblanc and R.B. Mann, Phys. Rev. {\bf D48}
(1993) 3606.
\item{[28]} R.~M. Wald, {\sl General Relativity} (University of Chicago
Press, Chicago, 1984).


\bye